\documentclass[onecolumn,12pt]{article}
\usepackage{graphicx}
\newcommand\simgt{\hspace{0.3em}\raisebox{0.4ex}{$>$}\hspace{-0.75em}\raisebox{-.7ex}{$\sim$}\hspace{0.3em}}

\begin{document}
\baselineskip 14pt

\title{Distortion of Magnetic Fields in a Starless Core III: \\
Polarization--Extinction Relationship in FeSt 1-457} 
\author{Ryo Kandori$^{1}$, Motohide Tamura$^{1,2,3}$, Tetsuya Nagata$^{4}$, Kohji Tomisaka$^{2}$,\\
Nobuhiko Kusakabe$^{3}$, Yasushi Nakajima$^{5}$, Jungmi Kwon$^{6}$, Takahiro Nagayama$^{7}$ \\
and \\
Ken'ichi Tatematsu$^{2}$\\
{\small 1. Department of Astronomy, The University of Tokyo, 7-3-1, Hongo, Bunkyo-ku, Tokyo, 113-0033, Japan}\\
{\small 2. National Astronomical Observatory of Japan, 2-21-1 Osawa, Mitaka, Tokyo 181-8588, Japan}\\
{\small 3. Astrobiology Center of NINS, 2-21-1, Osawa, Mitaka, Tokyo 181-8588, Japan}\\
{\small 4. Kyoto University, Kitashirakawa-Oiwake-cho, Sakyo-ku, Kyoto 606-8502, Japan}\\
{\small 5. Hitotsubashi University, 2-1 Naka, Kunitachi, Tokyo 186-8601, Japan}\\
{\small 6. Institute of Space and Astronautical Science, Japan Aerospace Exploration Agency,}\\
{\small 3-1-1 Yoshinodai, Chuo-ku, Sagamihara, Kanagawa 252-5210, Japan}\\
{\small 7. Kagoshima University, 1-21-35 Korimoto, Kagoshima 890-0065, Japan}\\
{\small e-mail: r.kandori@nao.ac.jp}}
\maketitle

\abstract{\bf 
The relationship between dust polarization and extinction was determined for the cold dense starless molecular cloud core FeSt 1-457 based on the background star polarimetry of dichroic extinction at near-infrared wavelengths. Owing to the known (three-dimensional) magnetic field structure, the observed polarizations from the core were corrected by considering (a) the subtraction of the ambient polarization component, (b) the depolarization effect of inclined distorted magnetic fields, and (c) the magnetic inclination angle of the core. After these corrections, a linear relationship between polarization and extinction was obtained for the core in the range up to $A_V \approx 20$ mag. The initial polarization vs. extinction diagram changed dramatically after the corrections of (a) to (c), with the correlation coefficient being refined from 0.71 to 0.79. These corrections should affect the theoretical interpretation of the observational data. The slope of the finally obtained polarization--extinction relationship is $P_H / E_{H-K_s} = 11.00 \pm 0.72$ $\%$ ${\rm mag}^{-1}$, which is close to the statistically estimated upper limit of the interstellar polarization efficiency (Jones 1989). This consistency suggests that the upper limit of interstellar polarization efficiency might be determined by the observational viewing angle toward polarized astronomical objects. 
}


\vspace*{0.3 cm}

\section{Introduction}
The relationship between dust dichroic polarization ($P$) and extinction ($A$) in molecular clouds and cores is important for two perspectives. First, the linearity in the $P$--$A$ relationship confirms the interpretation of the observed interstellar polarization angle, which is closely related to the direction of magnetic fields pervading the line-of-sight region. Second, theories of the alignment of dust grains can be restricted or confirmed by comparing them with observed $P$--$A$ relationships. \par
%
Goodman et al. (1995) proposed that the polarizing power of dust grains decreases inside cold dense dark molecular clouds. The severe decrease of the efficiency of polarization per unit extinction can be attributed to the physical and chemical properties of dust grains inside the clouds. Subsequent observational studies at optical wavelengths, Arce et al. (1998), revealed that there is a turnover in the $P$--$A$ relationship of stars toward the Taurus molecular cloud at $A_V = 1.3$ mag. The study suggests (1) stars that are background to the warm ISM show an increase in polarization with extinction, and (2) the polarization of stars that are background to cold dark clouds does not increase with extinction.
On the other hand, many polarimetric observations, particularly at near-infrared (NIR) wavelengths, show relatively linear $P$--$A$ relationships, with an observed polarization degree that increases with increasing dust extinction in the deep regions of a cloud interior, from massive star forming regions (e.g., NGC 2024: Kandori et al. 2007; M 42: Kusakabe et al. 2008) to low-mass star forming regions (e.g., Serpens cloud core: Sugitani et al. 2010). In studies of magnetic field strength as well as field geometry, the linearity of the $P$--$A$ relationship is crucial, because observational interpretations of magnetic fields fail if the detected polarization does not reflect the grain alignment inside the cold dense regions of molecular clouds. Accurately determining the $P$--$A$ relationship remains a problem for observational studies of dark clouds. \par
Determining the dust alignment mechanism in cold dense molecular clouds has been a long-standing problem in astrophysics (e.g., Davis \& Greenstein 1951; Andersson, Lazarian, \& Vaillancourt 2015 for review). One of the most promising mechanisms is grain alignment by radiative torque (e.g., Dolginov \& Mitrofanov 1976; Draine \& Weingartner 1996,1997; Lazarian \& Hoang 2007). Whittet et al. (2008) reported a declining $P$--$A$ relationship in the Taurus molecular cloud, and compared the observations with a simple radiative torque model using the data for background field stars and embedded young stars at NIR wavelengths. Their data sets, however, are dominated by dark cloud complexes and affected by the depolarization effects caused by different polarization angles toward the lines of sight and/or the distorted or complicated magnetic fields, which are difficult to correct for. Note that the situation is the same for the results of Arce et al. (1998) described above. It is thus important to determine the $P$--$A$ relationship for isolated starless globules, with a simple shape and components as well as a simple magnetic field geometry associated with the core. The choice of such objects minimizes the systematic effects of observed polarizations and enables the determination of accurate $P$--$A$ relationships in cold dense environments. \par
In the present study, an accurate $P$--$A$ relationship is determined toward the dense dark cloud core FeSt 1-457. The core is cataloged as a member of the dark cloud complex, the Pipe Nebula (as core \#109 in Alves, Lombardi, \& Lada 2007; Onishi et al. 1999; Muench et al. 2007), located in the direction of the Galactic center, at a distance of $130^{+24}_{-58}$ pc (Lombardi et al. 2006). Owing to the core's relatively isolated geometry, simple shape, and rich stellar field lying behind the core, a density structure study based on the Bonnor--Ebert model (Bonnor 1956; Ebert 1955) was conducted by measuring dust extinction at NIR wavelengths (Kandori et al. 2005). The physical properties of the core are well determined, namely a radius of $18500 \pm 1460$ AU (144$''$), mass of $3.55 \pm 0.75$ M$_{\odot}$, and central density of $3.5(\pm 0.99) \times 10^5$ cm$^{-3}$ at a distance of 130 pc (Kandori et al. 2005). FeSt 1-457 was observed in NIR polarimetry and an ``hourglass-shaped'' distorted magnetic field structure associated with the core was revealed (Kandori et al. 2017a, hereafter Paper I). A three dimensional (3D) magnetic field structure has been reported (Kandori et al. 2017b, hereafter Paper II), and the value of the line-of-sight magnetic inclination angle and the curvature of the 3D field were determined. \par
%
The polarizations from the core consist of four components/effects: (1) polarizations solely associated with the core, (2) superposition of ambient polarizations in the same line-of-sight but unrelated to the core, (3) depolarization caused by the distorted magnetic fields of the core, and (4) the line-of-sight inclination angle of the magnetic axis of the core. The latter three effects are measured and corrected to determine an accurate $P$--$A$ relationship for this well-defined starless core. 
%
Note that Alves et al. (2014, 2015) reported the $P$--$A$ relationship in the direction of FeSt 1-457 at NIR wavelengths. They determined the relationship without any corrections, and our result is consistent with their reports if we do not correct for effects (2), (3), and (4). 
%
%
%
\section{Data and Methods}
The NIR polarimetric data for determining the $P$--$A$ relationship of FeSt 1-457 is taken from Paper I. Observations were conducted using the $JHK$${}_{\rm s}$-simultaneous imaging camera SIRIUS (Nagayama et al. 2003) and its polarimetry mode SIRPOL (Kandori et al. 2006) on the IRSF 1.4-m telescope at the South African Astronomical Observatory (SAAO). SIRPOL can provide deep (18.6 mag in the $H$ band, $5\sigma $ for one-hour exposure) and wide-field ($7.\hspace{-3pt}'7 \times 7.\hspace{-3pt}'7$ with a scale of 0$.\hspace{-3pt}''$45 ${\rm pixel}^{-1}$) NIR polarimetric data.
\par
Figure 1 shows the polarization vectors of point sources toward FeSt 1-457, superimposed on the intensity image in the $H$ band. The core appears as a dark obscuration at the center of the image and the radius of the core (144$''$) determined by the density structure study (Kandori et al. 2005) is shown by a white circle. \par
The polarimetry data toward the core is the superposition of polarizations from both the core itself and the ambient medium which is unrelated to the core. The unrelated \lq \lq off-core'' polarization component was spatially fitted in $Q/I$ and $U/I$ in the $H$ band using the stars located outside of the core radius ($R>144''$). The distributions of the $Q/I$ and $U/I$ values are modeled as $f(x,y)=A + Bx + Cy$, where $x$ and $y$ are the pixel coordinates, and $A$, $B$, and $C$ are the parameters to be fitted. \par
The off-core regression vectors are subtracted from all the polarization vectors in order to construct the polarization vector map solely associated with the core. After subtracting unrelated polarization components, 185 stars located within the core radius were selected for determining the $P$--$A$ relationship (Figure 2). \par
In Figure 2, the magnetic field follows a distinct axisymmetric shape reminiscent of an hourglass. The white lines show the magnetic field direction inferred from the fitting with a parabolic function $y = g + gC{x^2}$, where $g$ specifies the magnetic field lines and $C$ determines the degree of curvature in the parabolic function. The results represent the first observational evidence of hourglass-shaped magnetic fields in a starless core (Paper I). \par
Assuming axial symmetry of the hourglass-shaped field, the 3D magnetic field structure can be modeled (Paper II). The 3D version of the simple function employed in Paper I, $z(r, \varphi, g) = g + gC{r}^{2}$ in cylindrical coordinates $(r, z, \varphi)$ is used to model the core magnetic fields, where 
$\varphi$ is the azimuth angle (measured in the plane perpendicular to $r$). In the function, the shape of the magnetic field lines is axially symmetric around the $r$ axis. The function $z(r, \varphi, g)$ thus has no dependence on $\varphi$. 
%
The polarization vector maps of the 3D parabolic model are shown in Figure 3 (taken from Paper II). The white lines show the polarization vectors, and the background color shows the polarization degree of the model core. The applied line-of-sight inclination angle is labeled in the upper-left corner of each panel. Comparing the model parabolic field with observations shows that the line-of-sight inclination angle of the magnetic field direction (magnetic axis) of the core is $45^{\circ} \pm 10^{\circ}$ with a magnetic curvature $C$ of $2.0 \times 10^{-4}$ ${\rm arcsec}^{-2}$ (Paper II). \par
The inclined distorted polarization fields in a model core produce a depolarization effect, particularly in the equatorial plane of the core. Since we virtually observe the model core with distorted magnetic fields from an inclined direction, the vectors located at the front and back sides of the core cross the polarization vector direction. In Figure 3, the depolarization regions can be seen as dark patches (apparent in the panels for $\theta_{\rm inc}=30^{\circ}$ and $\theta_{\rm inc}=15^{\circ}$), where the polarization degree is low compared with neighboring regions. The depolarization effect as well as inclination in the magnetic axis modify the observed polarization vectors from their original vector values. These effects can be corrected by using a 3D model and the best fit parameters for FeSt 1-457. 
%
%
\section{Polarization--Extinction Relationship}
Figure 4(a) shows the observed $P_H$ vs. $H-K_{s}$ relationship without any correction. It is clear from the figure that the polarization increases with increasing extinction up to $H-K_{s} \approx 0.9$ mag, and the relationship shows a relatively flat distribution in the region of $H-K_{s} \simgt 0.9$ mag. Such a ``kink'' is also seen in Alves et al. (2014, 2015), which seems to be evidence of a decrease of polarization efficiency inside the dense core. The location of the kink, $H-K_{s} = 0.9$ mag, corresponds to $A_V = 7.6$ mag, if we use the average $H-K_{s}$ color for the whole stars in the outside of the core radius, ${\left\langle{H-K}\right\rangle}_{\rm ref}=0.54$ mag, as a background stellar color, and use the reddening law of $A_V = 21.7 \times E_{H-K_s}$ by Nishiyama et al. (2008). The slope of the $P_H$ vs. $H-K_{s}$ relationship for whole data is $2.43(\pm 0.05)$ $\%$ ${\rm mag}^{-1}$. If we choose the data points of $H-K_{s} < 0.9$ mag, the slope of the relationship gets steeper as $2.92(\pm 0.15)$ $\%$ ${\rm mag}^{-1}$.
\par
The observed polarization is the superposition of the polarization from the core itself and ambient polarization which is unrelated to the core. A relatively linear $P_H$ vs. $H-K_{s}$ relationship is obtained after subtracting ambient polarization components, as shown in Figure 4(b) (see also Figure 6 of Paper I). The method of the subtraction of ambient polarization components is described in Section 2 (see also Paper I). The shape of the $P$--$A$ relationship changes, with disappearance of the kink and flattened feature in the region of $H-K_{s} \simgt 0.9$ mag. The gray plus symbols show ambient vectors ($R > 144''$ and $P/\delta P \ge 10$) after the subtraction analysis. The ambient vectors show no trend with suppression in polarization degrees. The subtraction analysis of ambient polarization thus seems successful. It is now clear that subtracting the background ambient polarization is crucial for a robust $P$--$A$ relationship. The slope of the relationship is $4.75(\pm 0.33)$ $\%$ ${\rm mag}^{-1}$, which is similar to the average interstellar polarization slope (Jones 1989). Comparing the slopes in Figures 4(a) and 4(b) data, the $P$--$A$ relationship gets steeper after the ambient subtraction. The shallow slope in Figure 4(a) data seems to be due to the contamination effect of ambient and core vector components that have different polarization angles.
\par
FeSt 1-457 is known to be accompanied by inclined distorted magnetic fields (Paper II), which causes depolarization effects inside the core. With a known 3D magnetic field structure, the depolarization correction factor can be calculated as shown in Figure 5. In Figure 5, severe depolarization regions around the equatorial plane clearly exist, resembling the model core simulation in Figure 3 (for the case of $45^{\circ}$). The depolarization correction factor was calculated by dividing the polarization degree distribution of the (best-fit) inclined distorted field model by that based on the inclined uniform field. \par
Figure 4(c) shows the depolarization corrected $P$--$A$ relationship obtained by dividing the Figure 4(b) relation by the depolarization correction factor shown in Figure 5. In Figure 4(c), the $P$--$A$ relationship becomes steeper than that in Figure 4(b), reflecting the correction. In the relationship, the relative location of the plot points moves slightly, because the depolarization correction factor is not spatially uniform. The slope of the relationship is $7.74(\pm 0.51)$ $\%$ ${\rm mag}^{-1}$. The correlation coefficient is 0.79, which is larger than those in Figures 4(a) and 4(b), with relationships of 0.71 and 0.76, respectively. It can be seen that the corrections based on the depolarization factor as well as the subtraction of ambient polarization components improves the tightness of the $P$--$A$ relationship. \par
Figure 4(d) shows the final correction, which is the relationship in Figure 4(c) divided by $\sin(45^{\circ})$, the correction using the magnetic inclination angle. The slope of the relationship is $11.00(\pm 0.72)$ $\%$ ${\rm mag}^{-1}$. The value of the slope is close to the statistically estimated upper limit of the interstellar polarization efficiency of $P_H / E_{H-K_s} \approx 14$ (Jones 1989). This consistency suggests that the upper limit of the interstellar polarization efficiency is determined by the observational viewing angle toward polarized astronomical objects, although it is possible that different dust properties in different clouds and/or different environments including various magnetic field strengths may play a role in determining of the upper limit of the interstellar polarization efficiency. Comparing Figures 4(a) and 4(d), the change is dramatic. After applying all the relevant corrections, we obtained the final, accurate relationship between the polarization and dust extinction. The relationship is linear up to $A_V \approx 20$ mag. 
\section{Implications for the Study of Dust}
Theories of the alignment of dust grains propose a decreasing polarization efficiency with increasing density or opacity (e.g., Lazarian et al. 1997). In contrast, our results show a clear linear $P$--$A$ relationship, and there is no turnover in the relationship in the cold and dense region in the starless core. Whittet et al. (2008) conducted comparison studies between the radiative torque theory and NIR polarimetric observations of stars toward the Taurus dark cloud complex. In their calculations, under realistic parameters based on observations, the polarization efficiency starts to drop at several to 10 mag in $A_V$. In contrast, for colder and denser conditions, FeSt 1-457 shows no drop in polarization efficiency. It may be difficult to explain the grain alignment in FeSt 1-457 using the radiative torque theory with general environment parameters and grain properties. \par
There are several mechanisms to increase the efficiency of radiative torque in cold dense starless cores. One is the strong interstellar radiation field (ISRF). Though FeSt 1-457 is in the Pipe Nebula dark cloud complex, it is geometrically possible for the core to be exposed toward the Galactic center and/or nearby bright stars, which can be a source of intense radiation field at NIR wavelengths. Note that the Pipe Nebula may be at the H II region shell swept by the nearby B-type star $\theta$ Ophiuchi (Gritschneder \& Lin 2012), and FeSt 1-457 might be illuminated by the NIR light from this bright star. Ascenso et al. (2013) and Forbrich et al. (2015) showed evidence of grain growth in FeSt 1-457 based on a comparison between mid-infrared (MIR) or sub-millimeter (submm) data and NIR extinction data. The efficiency of the radiative torque mechanism increases with increasing grain size and is optimized for a grain size comparable with the wavelength of incident light (Cho \& Lazarian 2005; Lazarian \& Hoang 2007). Moreover, in an environment with large grains, the penetration depth of ambient ISRF toward a dense cloud interior can be deeper (e.g., Strafella et al. 2001), which will also increase the efficiency of the radiative torque mechanism. If a clumpy structure exists in the core, this will also increase the ISRF penetration depth. Theoretical studies for comparison with the observations of FeSt 1-457 are necessary to explain the linearity in the polarization vs. extinction relationship at NIR wavelengths. 
\par


Note that the change of grain size can affect both polarization and extinction as well as their wavelength dependence. Cho \& Lazarian (2005) showed that the polarization vs. dust emission relationship changes depending on the size distribution of dust grains inside dense cloud based on the radiative torque theory. If large grains are abundant, the polarization degree toward dense interior of the core is enhanced due to high alignment efficiency of large grains. 
%
%
Since we obtained the linear relationship in $P$ and $A$, the most simple interpretation is that the polarizing power of dust grains does not change in the range from the core boundary to the deepest probing depth in our observations. The \lq \lq polarizing power'' can be related to the strength of radiation field and the size of dust grains. It is possible that the effect of the decrease of radiation strength and (increase of) the abundance of large grains from the core boundary to the central region may be balanced to create observed linear $P$--$A$ relationship. Further theoretical and observational studies are desirable for this topic. 
\par
Since the background starlight can get through the core material and show polarization, the local ISRF can also penetrate and contribute to the grain alignment to that depth. 
%
It is noteworthy that the probing depth of our polarization data is $A_V \approx 20$ mag that is less than the extinction toward the center of the core ($A_V \approx 40$ mag, Kandori et al. 2005). 
It is thus possible that the polarization efficiency toward the densest center may decrease from the value of outer regions. To confirm this, NIR imaging polarimetry of extreme depth toward the center of FeSt 1-457 using large telescopes is needed.
\\
\par
In the present study, the relationship between the dust dichroic polarization and extinction was determined for the dense starless molecular cloud core FeSt 1-457. Observed polarizations from the core were corrected by considering (a) subtraction of the ambient polarization component, (b) the depolarization effect of inclined distorted magnetic fields in the core, and (c) the magnetic inclination angle of the core. A linear relationship was obtained for the core in the range up to $A_V \approx 20$ mag. Further studies for theoretical explanation of dust alignment in FeSt 1-457 are desirable. For observations, caution must be paid in treating polarimetry data. In our case, the original observational polarization vs. extinction diagram changed dramatically after the corrections applied. These corrections should affect the theoretical interpretation of observational data. 
%
%
%
%
\subsection*{Acknowledgement}
We are grateful to the staff of SAAO for their kind help during the observations. We wish to thank Tetsuo Nishino, Chie Nagashima, and Noboru Ebizuka for their support in the development of SIRPOL, its calibration, and its stable operation with the IRSF telescope. The IRSF/SIRPOL project was initiated and supported by Nagoya University, National Astronomical Observatory of Japan, and the University of Tokyo in collaboration with the South African Astronomical Observatory under the financial support of Grants-in-Aid for Scientific Research on Priority Area (A) No. 10147207 and No. 10147214, and Grants-in-Aid No. 13573001 and No. 16340061 of the Ministry of Education, Culture, Sports, Science, and Technology of Japan. RK, MT, NK, and KT (Kohji Tomisaka) also acknowledge support by additional Grants-in-Aid Nos. 16077101, 16077204, 16340061, 21740147, 26800111, 16K13791, and 15K05032.

\clearpage 

\begin{figure}[t]  
\begin{center}
 \includegraphics[width=6.5 in]{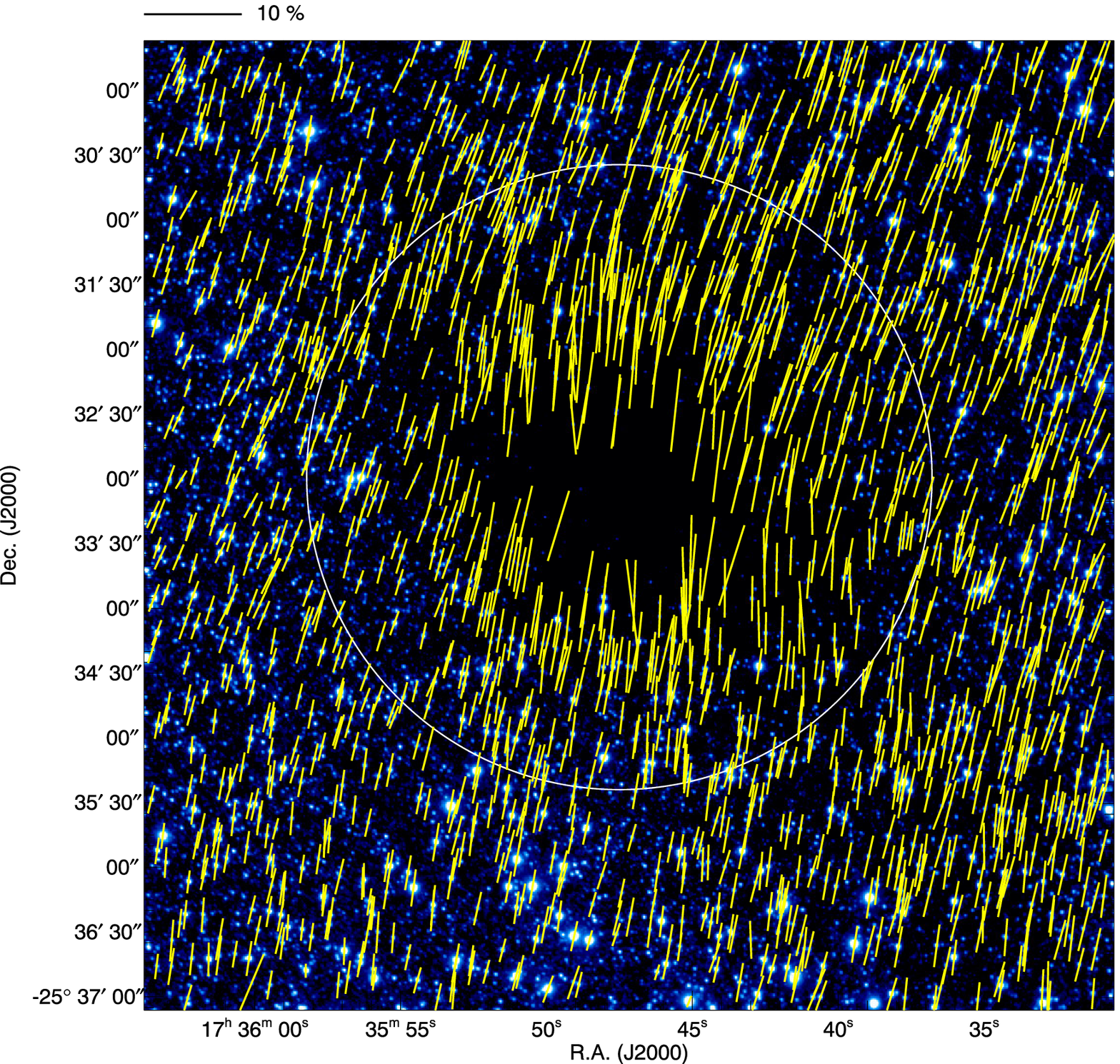}
 \caption{Polarization vectors of point sources superimposed on the intensity image in the $H$ band. The core radius (144$''$) is shown by the white circle. The scale of the 10$\%$ polarization degree is shown above the image. This figure is taken from Paper I. }
   \label{fig1}
\end{center}
\end{figure}

\clearpage 

\begin{figure}[t]  
\begin{center}
 \includegraphics[width=6.5 in]{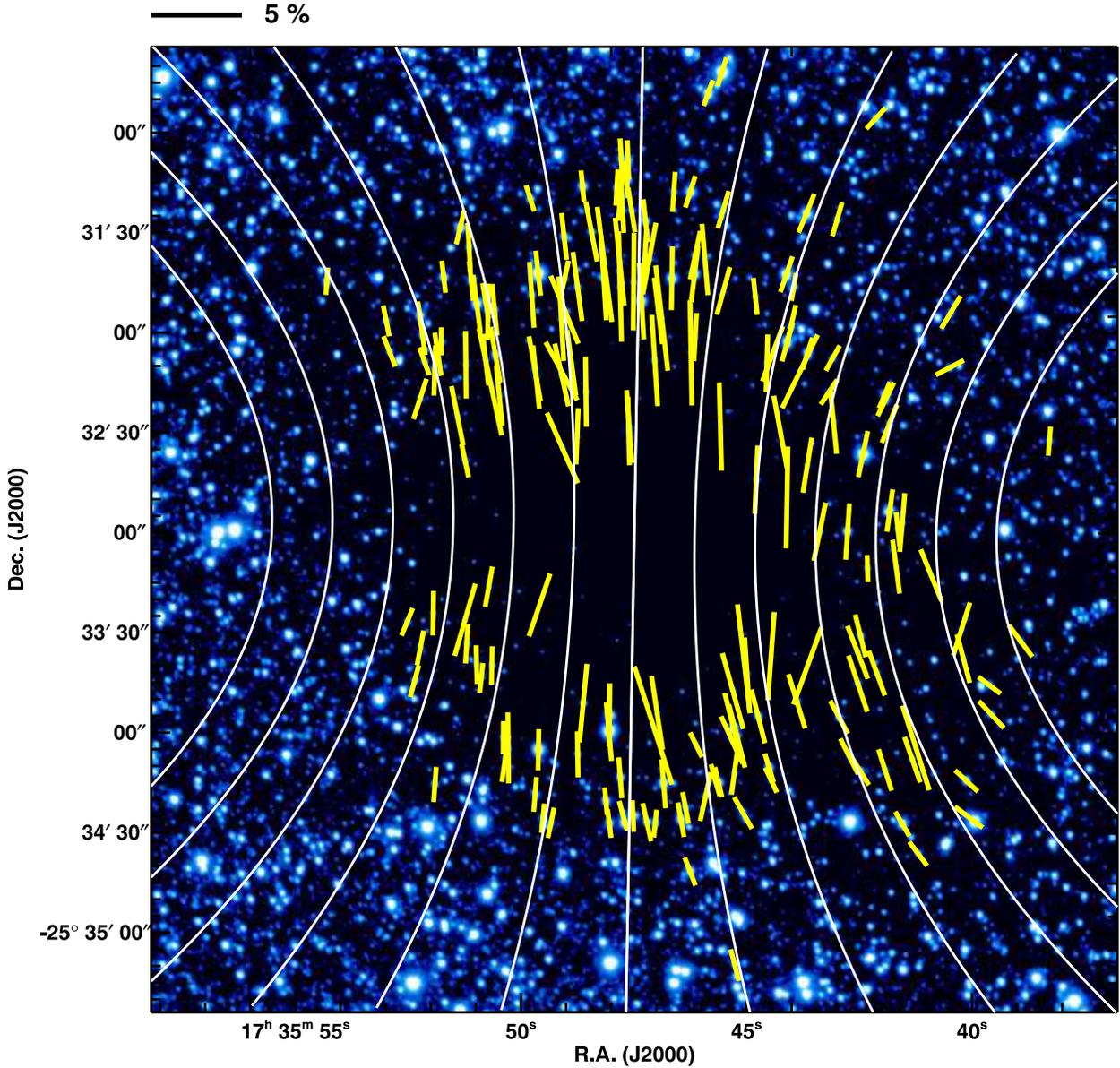}
 \caption{Polarization vectors of FeSt 1-457 after subtracting the ambient polarization component. The field of view is the same as the diameter of the core ($288''$ or 0.19 pc). The background image for this figure is a zoomed-in version of that used in Figure 1. The white lines indicate the magnetic field direction inferred from the fitting with a parabolic function of $y = g + gC{x^2}$, where $g$ specifies the magnetic field lines and $C$ determines the degree of curvature in the parabolic function. The scale of the 5$\%$ polarization degree is shown above the image. This figure is taken from Paper I.}
   \label{fig1}
\end{center}
\end{figure}

\clearpage 

\begin{figure}[t]  
\begin{center}
 \includegraphics[width=6.5 in]{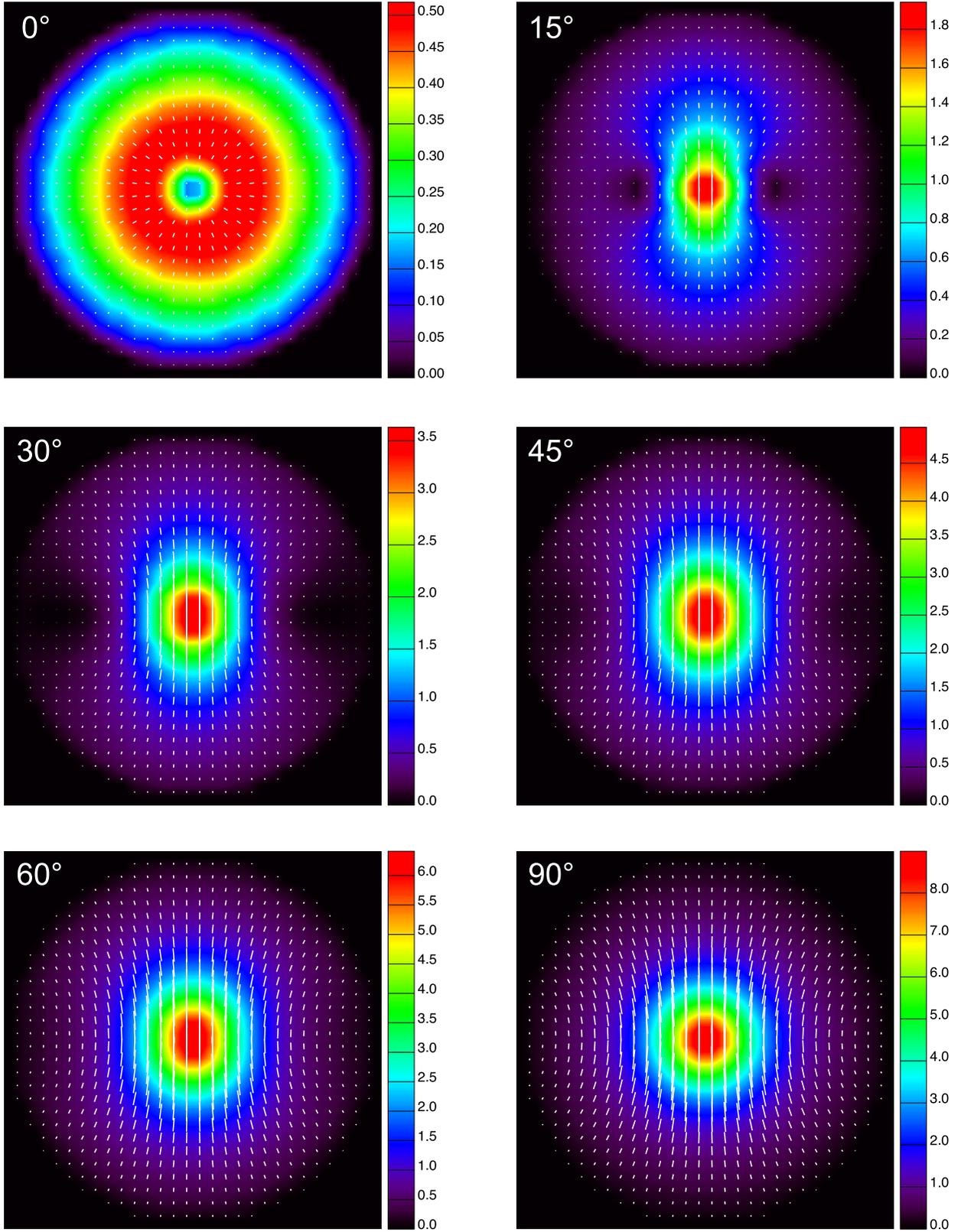}
 \caption{Polarization vector maps of the 3D parabolic model (white vectors). The background color shows the polarization degree. The applied line-of-sight inclination angle is labeled in the upper-left corner of each panel. The 3D magnetic curvature of the model is set to $C=2.5\times10^{-4}$ ${\rm arcsec}^{-2}$ for all the panels. This figure is taken from Paper II.}
   \label{fig1}
\end{center}
\end{figure}

\clearpage 

\begin{figure}[t]  
\begin{center}
 \includegraphics[width=6.5 in]{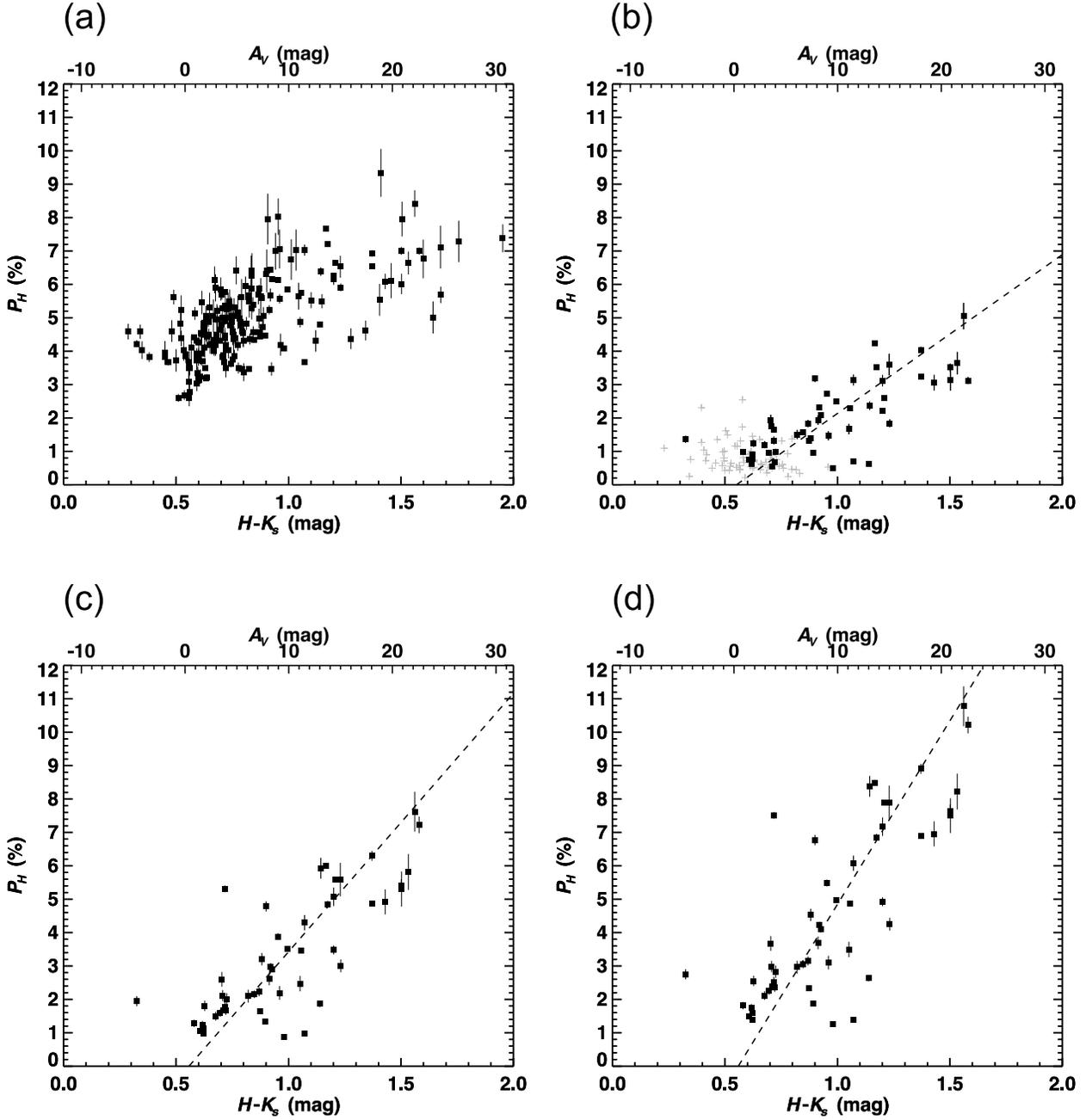}
 \caption{Relationship between polarization degree and $H-K_{\rm s}$ color toward background stars. Stars with $R \leq 144''$ and $P_H / \delta P_H \geq 10$ are plotted. (a) original $P$--$A$ relationship. (b) $P$--$A$ relationship after correcting for ambient polarization components. The gray plus symbols show the relationship for the stars located in the off-core region ($R > 144''$ and $P_H / \delta P_H \geq 10$). (c) $P$--$A$ relationship after correcting for ambient polarization components and the depolarization effect. (d) $P$--$A$ relationship after correcting for ambient polarization components, the depolarization effect, and the magnetic inclination angle.
}
   \label{fig1}
\end{center}
\end{figure}







\clearpage 

\begin{figure}[t]  
\begin{center}
 \includegraphics[width=6.5 in]{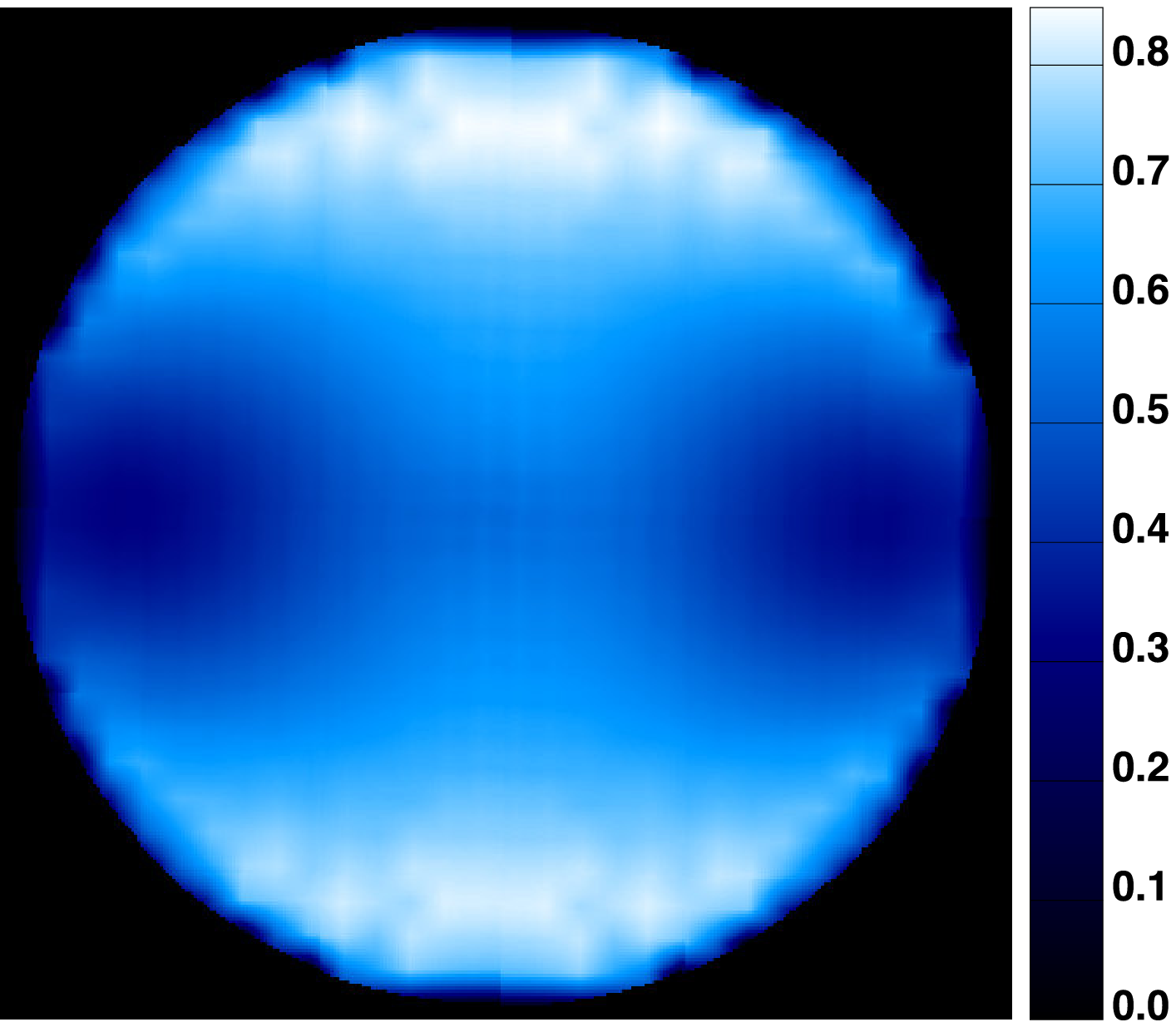}
 \caption{Distribution of depolarization correction factor. The field of view is the same as the diameter of the core ($288''$).}
   \label{fig1}
\end{center}
\end{figure}






\begin{thebibliography}{}
\bibitem{} Alves, F. O., Frau, P., Girart, J. M., et al., 2014, A\&AL, 569, 1
\bibitem{} Alves, F. O., Frau, P., Girart, J. M., 2015, A\&A, 574, 4
\bibitem{} Alves, J., Lombardi, M., \& Lada, C., 2007, A\&AL, 462, 17 
\bibitem{} Andersson, B.-G., Lazarian, A., \& Vaillancourt, J. E., 2015, ARA\&A, 53, 501 
\bibitem{} Arce, H. G., Goodman, A. A., Bastien, P., Manset, N., \& Sumner, M., 1998, ApJL, 499, 93 
\bibitem{} Ascenso, J., Lada, C. J., Alves, J., et al., 2013, A\&A, 549, 135 
\bibitem{} Bonnor, W. B., 1956, MNRAS, 116 351 
\bibitem{} Cho, J. \& Lazarian, A., 2005, ApJ, 631, 361 
\bibitem{} Davis, L. Jr., \& Greenstein, J. L., 1951, ApJ, 114, 206 
\bibitem{} Dolginov, A. Z., \& Mitrofanov, I. G., 1976, Ap\&SS, 43, 291 
\bibitem{} Draine, B. T., \& Weingartner, J. C., 1996, ApJ, 470, 551 
\bibitem{} Draine, B. T., \& Weingartner, J. C., 1997, ApJ, 480, 633 
\bibitem{} Ebert, R., 1955, ZA, 37, 217 
\bibitem{} Forbrich, J., Lada, C. J., Lombardi, M., Rom\'{a}n-Z\'{u}\~{n}iga, C., \& Alves, J., 2015, A\&A, 580, 114 
\bibitem{} Goodman, A. A., Jones, T. J., Lada, E. A., Myers, P. C., 1995, ApJ, 448, 748 
\bibitem{} Gritschneder, M., \& Lin, D. N. C., 2012, ApJ, 754, 13  
\bibitem{} Jones, T. J. 1989, ApJ, 346, 728 
\bibitem{} Kandori, R., Nakajima, Y., Tamura, M., et al., 2005, AJ, 130, 2166 
\bibitem{} Kandori, R., Kusakabe, N., Tamura, M., et al., 2006, Proc. SPIE, 6269, 159 
\bibitem{} Kandori, R., Tamura, M., Kusakabe, N., et al., 2007, PASJ, 59, 487  
\bibitem{} Kandori, R., Tamura, M., Kusakabe, N., et al., 2017a, ApJ, 845, 32 (Paper I) 
\bibitem{} Kandori, R., Tamura, M., Tomisaka, K., et al., 2017b, ApJ, 848, 110 (Paper II) 
\bibitem{} Kusakabe, N., Tamura, M., Kandori, R., et al., 2008, PASJ, 136, 621 
\bibitem{} Lazarian, A., 1997, ApJ, 490, 273 
\bibitem{} Lazarian, A. \& Hoang, T., 2007, MNRAS, 378, 910 
\bibitem{} Lombardi, M., Alves, J., \& Lada, C., 2006, A\&A, 454, 781 
\bibitem{} Muench, A. A., Lada, C. J., Rathborne, J. M., Alves, J. F., \& Lombardi, M., 2007, ApJ, 671, 1820
\bibitem{} Nagayama, T., Nagashima, C., Nakajima, Y., et al., 2003, Proc. SPIE, 4841, 459 
\bibitem{} Nishiyama, S., Nagata, T., Tamura, M., et al., 2008, ApJ, 680, 1174 
\bibitem{} Onishi, T., Kawamura, A., Abe, R., et al., 1999, PASJ, 51, 871
\bibitem{} Rathborne, J. M., Lada, C. J., Muench, A. A., Alves, J. F., \& Lombardi, M., 2008, ApJS, 174, 396 
\bibitem{} Strafella, F., Campeggio, L., Aiello, S., Cecchi-Pestellini, C., \& Pezzuto, S., 2001, ApJ, 558, 717 
\bibitem{} Sugitani, K., Nakamura, F., Tamura, M., et al., 2010, ApJ, 716, 299  
\bibitem{} Whittet, D. C. B., Hough, J. H., Lazarian, A., \& Hoang, T., 2008, ApJ, 674, 304 
\end{thebibliography}
\end{document}